\documentclass[letterpaper,aps,pre,twocolumn,showcaps,superscriptaddress,showpacs,amsmath,amssymb]{revtex4}
\usepackage{graphicx}

\newcommand{\dd}{\mathrm{d}}

\newcommand{\pd}[2]{\frac{\partial #1}{\partial #2}}

\newcommand{\mean}[1]{\langle #1 \rangle}
\newcommand{\Int}[1]{\int\dd #1\;}
\newcommand{\IInt}[3]{\int_{#2}^{#3}\dd #1\;}
\newcommand{\unit}[1]{\;\mathrm{#1}}

\renewcommand{\vec}[1]{\mathbf #1}

\newcommand{\al}{\alpha}

\newcommand{\eps}{\varepsilon}

\newcommand{\gam}{\gamma}

\newcommand{\X}{\Gamma}         
\newcommand{\x}{\vec r}         
\newcommand{\st}{_\mathrm{s}}   

\newcommand{\sm}{\mathcal L}
\newcommand{\ps}{\psi\st}

\newcommand{\ex}{\vec e_x}
\newcommand{\shr}{\gam}
\newcommand{\str}{\sigma_{xy}}  
\newcommand{\stra}{\tilde\sigma_{xy}}
\newcommand{\strt}{\bar\sigma_{xy}}

\newcommand{\F}{\mathbf{\mathcal F}}
\newcommand{\lb}{\lambda_\mathrm{B}}
\newcommand{\W}{\mathcal W}
\newcommand{\A}{\mathcal A}
\newcommand{\id}{\mathbf 1}

\begin{document}


\title{Extended Fluctuation-Dissipation Theorem for Soft Matter in Stationary
  Flow}

\author{Thomas Speck}
\altaffiliation[present address: ]{Department of Chemistry, University of
  California, Berkeley, CA 94720, USA}
\affiliation{{II.} Institut f\"ur Theoretische Physik,
  Universit\"at Stuttgart, Pfaffenwaldring 57, 70550 Stuttgart,
  Germany}
\author{Udo Seifert}
\affiliation{{II.} Institut f\"ur Theoretische Physik,
  Universit\"at Stuttgart, Pfaffenwaldring 57, 70550 Stuttgart,
  Germany}

\begin{abstract}
  For soft matter systems strongly driven by stationary flow, we discuss an
  extended fluctuation-dissipation theorem (FDT). Beyond the linear response
  regime, the FDT for the stress acquires an additional contribution involving
  the observable that is conjugate to the strain rate with respect to the
  dissipation function. This extended FDT is evaluated both analytically for
  Rouse polymers and in numerical simulations for colloidal suspensions. More
  generally, our results suggest an extension of Onsager's regression
  principle to nonequilibrium steady states.
\end{abstract}


\maketitle


\textit{Introduction. --} The fluctuation-dissipation theorem (FDT) holding
for all systems slightly perturbed around their equilibrium state is one of
the cornerstones of equilibrium statistical physics~\cite{kubo}. Specifically,
for a system coupled to a heat bath at temperature $T$ and setting Boltzmann's
constant to unity throughout the manuscript,
\begin{equation}
  \label{eq:fdt:orig}
  TR_{A,h}(t-\tau) = \mean{A(t)\dot B(\tau)} \equiv C_{A\dot B}(t-\tau)
\end{equation}
relates the response $R_{A,h}(t-\tau)\equiv\delta\mean{A(t)}/\delta h(\tau)$
of an observable $A$ to a small perturbation $h$ to equilibrium
correlations. Crucially, these correlations involve the same observable $A$
and the time-derivative $\dot B$ of another observable that is conjugated to
the perturbation $h$ in the sense that upon a perturbation the energy $U$ of
the system transforms as $U\mapsto U-Bh$. In a stationary state, due to
time-translational invariance both response and correlation function can only
depend on the difference $t-\tau>0$. The physical picture behind the FDT can
be expressed by Onsager's regression principle: the decay of fluctuations
created by a small external perturbation cannot be distinguished from the
decay of spontaneous thermal fluctuations.

Beyond the linear response regime, the FDT in the form~(\ref{eq:fdt:orig}) no
longer holds. However, more than 30 years ago, Agarwal noted that any
stationary Markov process obeys a generalized FDT~\cite{agar72}. This
generalized FDT is obtained through linear response theory of stochastic
processes where the probability distribution $\psi(t)$ obeys
$\partial_t\psi=\sm\psi$ with some time evolution operator
$\sm$~\cite{agar72,hang82}. In the presence of a perturbation, one can
decompose the evolution operator $\sm=\sm_0+h\delta\sm$ into an unperturbed
part $\sm_0$ and a perturbation operator $\delta\sm$. The stationary solution
of the unperturbed system obeys $\sm_0\ps=0$. The generalized FDT then has the
form of Eq.~(\ref{eq:fdt:orig}) with $\dot B$ replaced by $B^\ast\equiv
T\ps^{-1}\delta\sm\ps$. This form appears to be less useful than its
equilibrium counterpart because the new conjugate variable $B^\ast$ has no
independent physically significant meaning. In order to equip the involved
variables with such a physical interpretation, we can take the effect of the
driving into account as an additive excess function. The general structure of
such an extended FDT reads
\begin{equation}
  \label{eq:fdt:mod}
  TR_{A,h}(t) = \mean{A(t)[\dot B(0)-\bar B(0)]} \equiv C_{A\dot B}(t)
  - I_{A\bar B}(t).
\end{equation}
Interpreted in the spirit of Onsager's regression principle, the FDT now
states that the decay of forced fluctuations \textit{out of a nonequilibrium
  steady state} cannot be distinguished from the decay of spontaneous
fluctuations \textit{around} $\bar B$. Such a statement becomes physically
significant only if the variable $\bar B$ carries a transparent physical
meaning as an observable. One purpose of the present Rapid Communication is to
show that for shear flow driven soft matter systems governed by stochastic
dynamics, the extended FDT~(\ref{eq:fdt:mod}) acquires such a transparent
form: for a perturbation caused by a change of the strain rate the observable
$\bar B$ becomes the stress that is conjugated to the strain rate with respect
to the dissipation function, the mean of which is related to the total entropy
production. Such a characterization generalizes the identification of $\bar B$
as the local mean velocity in our previous study of a driven Langevin-type
dynamics with $h$ a small additional force~\cite{spec06}. A change of the
frame of reference from the laboratory frame to the frame moving with this
local mean velocity then restores the equilibrium form of the
FDT~\cite{spec06,chet08}. The extended FDT in its integrated form leads to an
experimentally tested generalized Einstein relation~\cite{blic07}.

Our approach is complementary to the strategy of introducing an effective
temperature to restore the equilibrium form~(\ref{eq:fdt:orig}) of the FDT
even in nonequilibrium. The latter approach has been developed over the last
decade for systems with a small heat flow into the reservoir corresponding to
a small entropy production rate~\cite{cugl97a,barr00,bert02,haxt07}. With such
an effective temperature concepts from equilibrium statistical mechanics could
be applied to driven systems even though a full microscopic understanding on
the range of validity of this concept does not seem to have been reached yet.

For the restricted but paradigmatic class of shear driven systems on which we
will focus, quantitative progress has been achieved using the framework of
mode-coupling theory. This includes the constitutive equation~\cite{brad08}
using integration through transients~\cite{fuch05} and an FDT for the
diffusion of a tagged particle~\cite{szam04}. Invariant
quantities~\cite{baul08} constitute an exact result for systems driven through
the boundaries with an unchanged bulk Hamiltonian, whereas in this Rapid
Communication the system is driven through an imposed external flow.


\textit{Soft matter under shear. --} We consider soft matter systems such as
colloidal suspensions or polymers that can be described as $N$ interacting
Brownian particles. The system is driven into a nonequilibrium steady state by
shearing with a strain rate $\shr$, resulting in an imposed flow profile $\vec
u(\x)=\shr y\ex$ of the solvent with unit vector $\ex$. Picking out the $i$th
particle, the force exerted by all other particles is $\vec
F_i\equiv-\nabla_iU$, where the potential energy
$U(\X)=\sum_{(ij)}u(|\x_{ij}|)$ is given by the sum over all pairs $(ij)$ with
$\x_{ij}\equiv\x_i-\x_j$. The particles interact via an isotropic pair
potential $u(r)$ and the set of particle positions is denoted as
$\X\equiv\{\x_1,\dots,\x_N\}$~\footnote{Even though hydrodynamic interactions
  can easily be included in our formalism through symmetric mobility tensors
  $\mu_{ij}(\X)$, here, for the sake of brevity, we use
  $\mu_{ij}=\mu_0\delta_{ij}\id$ neglecting any hydrodynamic interaction
  between the $N$ particles. The bare mobility $\mu_0$ and the short-time
  diffusion coefficient $D_0$ are connected through the Einstein relation,
  $D_0=\mu_0T$.}.

The response of an observable $A(\X)$ to a small, time-dependent variation of
the strain rate is $R_{A,\shr}(t-\tau;\shr)$, where the dependence on $\shr$
emphasizes that such a response can be defined for any steady state, not only
for $\shr=0$ corresponding to equilibrium. The mean
$\mean{A(t)}\equiv\Int{\X}A(\X)\psi(\X,t)$ involves the time-dependent
probability distribution $\psi(\X,t)$. The perturbation operator with respect
to a small change of the strain rate is $\delta\sm=-\sum_iy_i\partial/\partial
x_i$. In the linear response regime, the FDT
\begin{equation}
  \label{eq:fdt}
  T R_{A,\shr}(t-\tau;0) = \mean{A(t)\str(\tau)}
\end{equation}
relates the response to a correlation function involving the stress
\begin{equation}
  \label{eq:stress}
  \dot B = \str = \sum_{(ij)}\frac{x_{ij}y_{ij}}{|\x_{ij}|}
  \pd{u(|\x_{ij}|)}{r} = -\sum_{i=1}^N y_i\ex\cdot\vec F_i
\end{equation}
due to particle interactions. The resulting generalized
fluctuation-dissipation relation reads
\begin{equation}
  \label{eq:fdr}
  TR_{A,\shr}(t-\tau;\shr) = \mean{A(t)\str^\ast(\tau)}
\end{equation}
with conjugate stochastic variable
\begin{equation}
  \label{eq:stress:conj}
  \str^\ast \equiv T\ps^{-1}\delta\sm\ps
  = -\sum_{i=1}^Ny_i\ex\cdot T\nabla_i\ln\ps.
\end{equation}
Applying the generalized Onsager principle by following
Eq.~(\ref{eq:fdt:mod}), we split $\str^\ast=\str-\strt$ into the
stress~(\ref{eq:stress}) and
\begin{equation}
  \label{eq:stress:loc}
  \strt \equiv -\sum_{i=1}^N y_i\ex\cdot\F_i
\end{equation}
which involves the thermodynamic force $\F_i\equiv-\nabla_i[U+T\ln\ps]$. Both
stresses have the same mean $\mean{\str}=\mean{\strt}$. In equilibrium,
$\strt=0$ vanishes and hence Eq.~\eqref{eq:fdr} reduces to Eq.~\eqref{eq:fdt}
as expected.

We now provide for $\strt$ a clear physical meaning connecting it to the
entropy production caused by the external flow $\vec u(\x)$. For overdamped
dynamics, the dissipation function~\cite{doiedwards}
\begin{equation}
  \label{eq:diss}
  \W(\X,\{\vec v_i\};\shr) = \frac{1}{2\mu_0}\sum_{i=1}^N
  [\vec v_i-\vec u(\x_i)]^2
\end{equation}
is related to the mean total entropy production rate through $T\mean{\dot
  s_\mathrm{tot}}=2\mean{\W}$~\cite{spec08}. In a nonequilibrium steady state,
the sum $\mean{\W}+\dot\A$ of mean dissipation function and time-derivative of
the ``dynamical free energy'' $\A\equiv\Int{\X}\psi[U+T\ln\psi]$ attains a
minimum with respect to the local mean velocities $\{\vec v_i\}$, which then
obey $\vec v_i=\vec u(\x_i)+\mu_0\F_i$. A variation of $\W$ with respect to
the strain rate leads to
\begin{equation*}
  \pd{\W}{\shr} = -\frac{1}{\mu_0}\sum_{i=1}^N[\vec v_i-\vec
  u(\x_i)]\cdot\pd{\vec u(\x_i)}{\shr} = \strt.
\end{equation*}
In this sense, $\strt$ is the variable conjugate to the strain rate $\shr$
with respect to the dissipation function in analogy to $B$ being the variable
conjugate to $h$ with respect to the energy.

In the remainder of this Rapid Communication, we concentrate on response and
correlation functions of $A=\dot B=\str-\mean{\str}$ and $\bar
B=\strt-\mean{\str}$, e.g., $C(t)=\mean{\str(t)\str(0)}-\mean{\str}^2$. To
ease the notation we drop the subscripts denoting the observables. We first
consider a Rouse polymer which allows for analytic expressions and then turn
to numerical results for a colloidal suspension with nontrivial interactions.


\textit{Rouse polymer. --} Analytic expressions for correlation and response
functions can be obtained for systems with quadratic interaction energies of
the form $U(\X)=\frac{1}{2}\sum_\al k_\al\vec q_\al^2$ with amplitudes $\{\vec
q_\al\}$ of normal modes. The stress through interactions is $\str=\sum_\al
k_\al x_\al y_\al$. Inserting the explicit form of $\ps$,
Eq.~(\ref{eq:stress:loc}) becomes
\begin{equation}
  \label{eq:strt:rouse}
  \strt = \sum_\al \frac{k_\al}{1+\kappa_\al^2}\left( \kappa_\al^2 x_\al y_\al
    + \kappa_\al y_\al^2 \right).
\end{equation}
A straightforward calculation~\footnote{See supplementary information.} based
on the Smoluchowski operator leads to a closed equation of motion for the
correlation function
\begin{equation}
  \label{eq:C:rouse}
  C(t) = T^2 \sum_\al[1+\kappa_\al^2(3+2t/\tau_\al)]e^{-t/\tau_\al},
\end{equation}
where $\kappa_\al\equiv\shr\tau_\al$ and $\tau_\al\equiv(2\mu_0k_\al)^{-1}$ is
half the relaxation time of the corresponding mode. The response function
$R(t)=T\sum_\al e^{-t/\tau_\al}$ is independent of the driving. The excess
$I=C-TR\propto\shr^2$ is a quadratic function of the shear rate. While such
quadratic behavior is expected universally at small shear rates, for the Rouse
polymer its persistence for large $\shr$ depends on the Gaussian form of
$\ps$.

\begin{figure}[t]
  \centering
  \includegraphics[width=\linewidth]{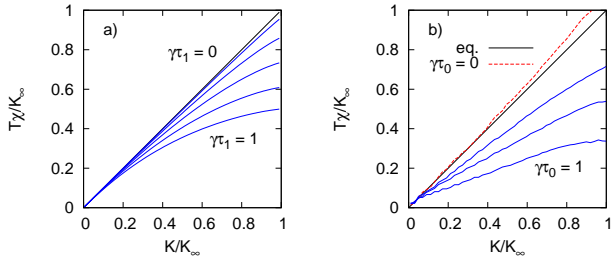}
  \caption{(color online) Integrated response $\chi$ \textit{vs}. integrated
    correlation function $K$ for a) Rouse polymer and b) a colloidal
    suspension (see main text). Both functions have been normalized by
    $K_\infty\equiv K(t\rightarrow\infty)$. The straight solid line
    corresponds to the equilibrium FDT~(\ref{eq:fdt}).}
  \label{fig:fdr}
\end{figure}

To obtain the universal expressions for a large number of modes, we replace
the summation by an integration, $\sum_\al\mapsto\int_1^\infty\dd\al$, and set
the relaxation times $\tau_\al=\tau_1/\al^2$, where the time scale is
determined by the fundamental relaxation time $\tau_1$. In addition, we
consider integrated response $\chi(t)\equiv\IInt{\tau}{0}{t}R(\tau)$ and
correlation $K(t)\equiv\IInt{\tau}{0}{t}C(\tau)$. In Fig.~\ref{fig:fdr}a) the
normalized integrated response $\chi(K)$ is plotted against the value of the
correlation function (parametrized by time). If the equilibrium
FDT~(\ref{eq:fdt}) holds then this curve is a straight line. For increasing
strain rates the deviation, and therefore the excess, increases. Moreover, for
the Rouse polymer no regimes with constant slope corresponding to an effective
temperature can be discerned.

\textit{Colloidal suspension. --} We consider $N$ colloidal particles with
diameter $a$ suspended in a fluid. We assume the particles to interact through
a repulsive, screened Coulomb pair potential
\begin{equation}
  \label{eq:pot}
  u(r) = TZ^2\lb\frac{e^{\kappa a}}{(1+\kappa a/2)^2}\frac{e^{-\kappa r}}{r},
\end{equation}
where $\lb\simeq7\unit{nm}$ is the Bjerrum length in water at room
temperature, $Z$ is the effective surface charge, and $\kappa^{-1}$ is the
screening length.

\begin{figure}[t]
  \centering
  \includegraphics[width=\linewidth]{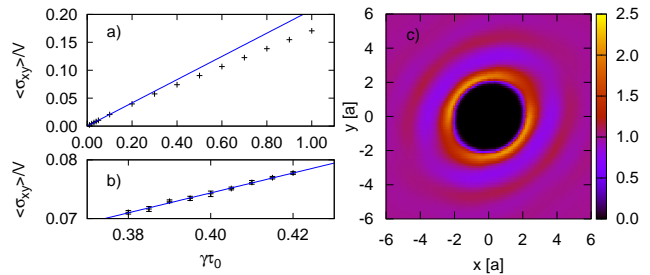}
  \caption{(color online) a)~Mean stress $\mean{\str}$ of a colloidal
    suspension \textit{vs}. the strain rate $\shr$. The solid line indicates
    the linear response regime. b)~Mean stress in the vicinity of
    $\shr\tau_0=0.4$. The solid line is a fit indicating the linear response
    of the nonequilibrium steady state. The error bars have been obtained as
    standard deviation by splitting the trajectory into eight
    segments. c)~Pair distribution function $g(\x)$ for $\shr\tau_0=1$ in the
    $xy$-plane with $z=0$. For parameters, see main text.}
  \label{fig:mean}
\end{figure}

Response and correlation functions are obtained through simulations of a
sheared dilute colloidal suspension in a cubic box with side length $L=25a$
and volume $V=L^3$. We are interested in the bulk behavior and therefore we
employ periodic Lees-Edwards boundary conditions in the simulation. The
particle number is $N=1000$ corresponding to a volume fraction of
$\phi\simeq0.034$. The screening length is set to $\kappa^{-1}=0.15a$ and the
effective surface charge is $Z=12000$. The natural time scale $\tau_0\equiv
a^2/D_0$ is set by the time a particle needs to diffuse a distance equal to
its diameter. To make contact with physical units, we choose $a=1\unit{\mu
  m}$. In Fig.~\ref{fig:mean}a), the mean stress is shown for different
P\'eclet numbers $\shr\tau_0$. The straight line indicates the linear response
behavior. The deviation of the mean stress from this line for large strain
rates corresponds to shear thinning of the suspension~\cite{hunter}. Mean
values and correlation functions are obtained from single runs with constant
strain rate. The response function is determined as
$R(t)=\partial_t\mean{\str(t)}/\eps$ after a jump $\shr\mapsto\shr+\eps$ of
the strain rate. Fig.~\ref{fig:mean}b) shows the linear response of the
nonequilibrium steady state corresponding to $\shr\tau_0=0.4$. In the
simulation, we have chosen a step of $\eps\tau_0=0.01$.

\begin{figure*}[t]
  \centering
  \includegraphics[width=\linewidth]{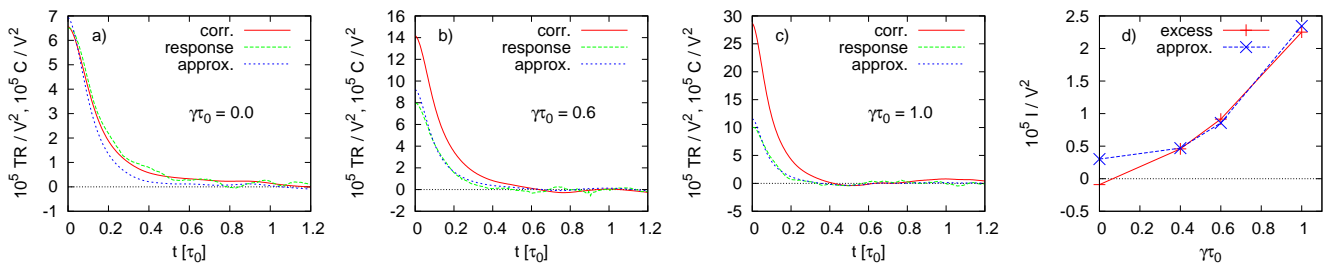}
  \caption{(color online) Correlation function $C(t)$, response function
    $R(t)$, and approximated response function $\tilde R(t)$ over time for a)
    equilibrium and b-c) two different nonequilibrium steady
    states. d)~Comparison of integrated excess (through $I=C-TR$) and
    integrated approximated excess [using Eq.~(\ref{eq:stress:app})]
    \textit{vs}. strain rate.}
  \label{fig:corr}
\end{figure*}

The strong violation of the FDT can been seen in Fig.~\ref{fig:corr}a-c),
where response and correlation functions are plotted as functions of time for
equilibrium ($\shr=0$) and two driven nonequilibrium steady states. For
increasing strain rate, the deviation of the correlation function $C(t)$ from
the response function $R(t)$ becomes larger, too. In Fig.~\ref{fig:fdr}b), the
normalized integrated response $\chi(K)$ is shown as function of the
integrated correlations. The small deviation between response and correlation
function in Fig.~\ref{fig:corr}a) is responsible for the fact that in
Fig.~\ref{fig:fdr}b) the equilibrium curve for $\shr=0$ lies slightly above
the expected straight slope. Overall, the structure is comparable to the Rouse
polymer in Fig.~\ref{fig:fdr}a). Following Ref.~\cite{cugl97a}, one might even
be tempted to identify a linear slope in the intermediate range corresponding
to an effective temperature even though this is not the focus of the present
work.

\textit{Approximate excess function. --} So far, we have obtained the excess
as difference $I=C-TR$. In principle, we need the complete distribution
$\ps(\X)$ to determine $I$ independently using Eq.~(\ref{eq:stress:loc}). Such
complete information is, however, neither available experimentally nor in
computer simulations. Hence, approximate schemes will become important in
future applications of the extended FDT. As a first step, we discuss here an
approximation to $\strt$. Since the suspension is homogeneous, the one-point
density $\rho^{(1)}(\x_1)=\rho=N/V$ is constant and the two-point density
\begin{equation*}
  \rho^{(2)}(\x_1,\x_2) = N(N-1)\int\dd\x_3\cdots\dd\x_N\;\ps(\X) 
  \equiv \rho^2 g(\x)
\end{equation*}
becomes a function of the displacement $\x\equiv \x_1-\x_2$ only. The factor
$N(N-1)$ accounts for the possible permutations of the identical particles.
The pair distribution $g(\x)$ as obtained from the simulation for the
parameters introduced above is shown in Fig.~\ref{fig:mean}c) (for hard
spheres, cf. Ref.~\cite{berg02}).

We approximate the stationary distribution as
$\tilde\ps(\X)=\exp\{-\alpha\sum_{(ij)}w(\x_{ij})/T\}$ with potential of mean
force $w(\x)\equiv-T\ln g(\x)$~\cite{hunter}. This approximation effectively
factorizes the probability distribution by using the correct pair correlations
and neglecting correlations between three and more particles. It is motivated
by the fact that the stress is determined by pair interactions only. In the
parameter range we have studied we found $\alpha\simeq0.5$. Inserting the pair
approximation into Eq.~(\ref{eq:stress:conj}) leads to
\begin{equation}
  \label{eq:stress:app}
  \stra^\ast 
  = \alpha \sum_{(ij)} y_{ij}\pd{w(\x_{ij})}{x_{ij}} + \mathrm{const}.
\end{equation}
The constant offset is adjusted for every strain rate such that
$\mean{\stra^\ast}=0$. We also employ a cut-off taking into account only
neighboring particles within the first shell. Using this approximation, we can
then both calculate the response function through~(\ref{eq:fdr}) and access
the stress~(\ref{eq:stress:loc}) in the simulation. In Fig.~\ref{fig:corr}a-c)
the approximated response function $\tilde R(t)$ is shown together with $R(t)$
and the correlation functions. In Fig.~\ref{fig:corr}d), both the integrated
excess $\IInt{\tau}{0}{\infty}I(\tau)$ as well as the integrated approximate
excess based on Eq.~(\ref{eq:stress:app}) are shown for different strain
rates. For moderate to large strain rates, the pair approximation works quite
well. In the limit of vanishing strain rate, even though the volume fraction
of the colloidal particles is low, the potential of mean force still deviates
from the pair potential resulting in a break-down of this type of
approximation for equilibrium.


\textit{Concluding perspective. --} For strongly driven soft matter systems,
we have discussed an extended FDT. Beyond the analytical and numerical data
for two case studies, our general insight is twofold. First, beyond the linear
response regime, the FDT acquires an additive contribution which involves the
stress that in the dissipation function is conjugate to the strain rate. This
result suggests more generally that the nonequilibrium form of the FDT
involves the observable that is conjugate to the perturbation in the
dissipation function. Such a scheme could be analogous to the pairing of
observables conjugate with respect to energy in the equilibrium form of the
FDT. Second, the additive contribution allows an interpretation in the spirit
of Onsager's regression principle: The decay of a spontaneous fluctuation
\textit{around a nonequilibrium steady state} cannot be distinguished from the
decay of a fluctuation forced by a small external perturbation. Whether these
observations can be generalized to an even larger class of nonequilibrium
systems remains to be investigated both by further case studies and, more
ambitiously, by an attempt to formulate a more formal theory for
nonequilibrium steady states along these lines.

We acknowledge financial support by Deutsche Forschungsgemeinschaft through
SE~1119/3. While finishing this manuscript, TS was funded by the Helios Solar
Energy Research Center which is supported by the Director, Office of Science,
Office of Basic Energy Sciences of the U.S. Department of Energy under
Contract No.~DE-AC02-05CH11231.


\end{document}